\makeatletter\def\input@path{
	{./graphics/}
}\makeatother%
\documentclass[twocolumn,preprintnumbers,pra,superscriptaddress]{revtex4}%smaller,showpacs,showkeys

\usepackage{amssymb,amsmath,graphicx,dcolumn,bm,float} 
\usepackage[utf8]{inputenc}
\usepackage[T1]{fontenc}
\usepackage{xcolor}

\begin{document}

%\preprint{Physical Review A}
\title{Sensor sensitivity based on synthetic magnetism exceptional points}
\author{S. R. Mbokop Tchounda}
\email{rolande.mbokop@facsciences-uy1.cm}
\affiliation{Department of Physics, Faculty of Science, University of Yaounde I, P.O. Box 812, Yaounde, Cameroon}

\author{P. Djorwé}
\email{djorwepp@gmail.com}
\affiliation{Department of Physics, Faculty of Science, 
University of Ngaoundere, P.O. Box 454, Ngaoundere, Cameroon}
\affiliation{Stellenbosch Institute for Advanced Study (STIAS), Wallenberg Research Centre at Stellenbosch University, Stellenbosch 7600, South Africa}

\author{S. G. Nana Engo}
\email{serge.nana-engo@facsciences-uy1.cm}
\affiliation{Department of Physics, Faculty of Science, University of Yaounde I, P.O. Box 812, Yaounde, Cameroon}

\author{B. Djafari-Rouhani}
\email{bahram.djafari-rouhani@univ-lille.fr}
\affiliation{Institut d’Electronique, de Microélectronique et Nanotechnologie, UMR CNRS 8520 Université de Lille, Faculté des sciences et technologies, 59652 Villeneuve d’Ascq France}

\begin{abstract}
An efficient mass sensor based on exceptional points (EPs), engineered under synthetic magnetism requirement, is proposed. The benchmark system consists of an electromechanical (optomechanical) system where an electric (optical) field is driving two mechanical resonators which are mechanically coupled through a phase-dependent phonon-hopping. This phase induces series of EPs once it matches the condition of $\frac{\pi}{2}(2n +1)$. For any perturbation of the system, the phase-matched condition is no longer satisfied and this lifts the EP-degeneracies leading to a frequency splitting that scales as the square root of the perturbation strength, resulting in a giant sensitivity-factor enhancement. Owing to the set of EPs, our proposal allows multiple sensing scheme and performs better than its anti-PT-symmetric sensor counterpart. This work sheds light on new platforms that can be used for mass sensing purposes, opening up new opportunities in nanoparticles or pollutants detection, and to water treatment.
\end{abstract}

\maketitle

\date{\today}

%\preprint{APS/123-QED}

%
\section{Introduction} \label{Intro}
Sensing is one of the most important technological application of electromagnetism. Sensor applications impact our daily life and they cover versatile platform including viruses or nanoparticles-like mass detection \cite{Marquez_2017},  environmental pollution \cite{Ahmed_2021}, temperature sensor \cite{Zaremanesh_2021}, biosensors \cite{Quotane_2022}, salinity detection and water treatment \cite{Amiri_2018,Lucklum_2021,Lucklum_2009}. Owing to these practical implications of sensing, it is crucial to seek for a sensitivity improvement. A lot of attention has been recently paid to exceptional points (EPs), which are non-Hermitian singularities where the eigenvalues and their corresponding eigenstates coalesce. These interests have been raised due to the counter-intuitive features and intriguing effects happening at the vicinity of EPs. Among them are stopping light \cite{Goldzak_2018}, loss-induced suppression and revival of lasing, pump-induced lasing death,  unidirectional invisibility \cite{Peng_2014}, collective phenomena \cite{Djorwe_2018,Djorwe_2020}, and sensors.
The interest to non-Hermitian sensors lies on the topological feature at the EP which induces sensitivity enhancement resulting from the frequency splitting that exhibits a square-root dependency on a small perturbation \cite{Miri_2019}. This exceptional point sensitivity has been investigated in different fields including optics \cite{Hodaei_2017,Chen_2017}, plasmonic \cite{Park.2020}, electronics \cite{Song_2021,Dong_2019,Cao_2022,Chen_2019}, mechanics \cite{Rosa_2021}, and optomechanics \cite{Djorwe_2019}. Experimental comparison between optical sensor based-EP and on Diabolic Point (DP), that is the conventional hermitian degeneracy, has been put forwards in \cite{Chen_2017,Park.2020}, while a sensitivity enhancement at the higher order EP has been demonstrated in \cite{Hodaei_2017}. A practical electronics EP-microsensor has been successfully implemented on a rat skin in \cite{Dong_2019}, and a better wireless sensing response has been recorded and compared to the conventional approach based on DP. Despite the different aspects of these systems, what is common to them is the EP engineering approach which for most of them request amplification mechanism (gain) which is detrimental for sensing at the quantum level. Hencefore, engineering EPs without gain requirement has become a scrucial key to better sensing. To overcome this limitation, Anti Parity Time (APT) symmetry \cite{Li_2020} and artificial/synthetic magnetic field have been put forwards as mechanisms to engineer EPs in passive systems. Different approaches to create artificial magnetic field have been proposed for specific purposes such as non-reciprocal and topological bosonic transport \cite{Mathew_2020,Fang_2012,Fang_2017,Schmidt.2015}, heating-resistant ground-state cooling \cite{Jiang_2021}, and dark-mode breaking induced noise-tolerant entanglement \cite{Lai_2022}.

In this work, we propose a sensitivity enhancement of sensor based-EP that is engineered via synthetic magnetism. Our benchmark system consists of two mechanical resonators mechanically coupled, and driven by a common optical or electrical signal. This system supports a phase-dependent phonon-hopping interaction (e.g. a phase modulation of the mechanical coupling strength), which is introduced to create synthetic gauge fields. In contrast to the aforementioned sensors, the sensors based on synthetic magnetism do not require an amplification mechanism which is  usually  a tricky task in experiment. Beside on getting rid of this amplification process, we have shown that this new class of sensors perform better than the passive APT-symmetric sensors. Therefore, the presented approach of sensing gives new insights to improve sensitivity and opens up new opportunities for sensor applications.

This work is organized as follows. In Sec. \ref{sec:MoEq}, the model and its dynamical equations up to the EP features are described.  Section \ref{sec:Sen} is devoted to the sensitivity based on synthetic magnetism feature, and Sec. \ref{sec:Concl} concludes the work. 

\section{Modelling and dynamical equations} \label{sec:MoEq}

\begin{figure}[tbh]
  \begin{center}
  \resizebox{0.45\textwidth}{!}{
  \includegraphics{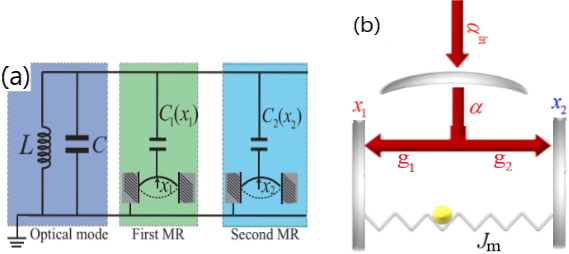}}
  \end{center}
  \caption{(a) Electromechanical system where two mechanically coupled mechanical resonators are driven by a common electrical field from a $LC-$oscillator. (b) An optomechanical equivalent system as in (a), where the yellow dot represents a small perturbation. The mechanical coupling rate $J_m$ is $\theta$ phase-dependent.}
  \label{fig:Fig1}
  \end{figure}

Our benchmark system is an electromechanical system where two mechanically coupled mechanical resonators are driven by a common electrical field generated from a $LC-$ oscillator as shown in Fig. \ref{fig:Fig1}a. The mechanical coupling rate $J_m$ between the two mechanical resonators is  modulated through a phase $\theta$, which is introduced to create synthetic gauge fields \cite{Lai_2022}. The same circuit can be modelled in an optomechanical system \cite{Mathew_2020} as shown in Fig. \ref{fig:Fig1}b.
In the rotating frame of the driving fields ($\omega_p$), the
Hamiltonian ($\hslash=1$) describing this system is
\begin{equation}\label{Hamil1}
H=H_{OM}+H_{int}+H_{drive}+H_{\kappa}+H_{\gamma},
\end{equation}  
where  
\begin{align}\label{Hamil2}
H_{OM}&=-\Delta a^{\dag}a+\sum_{j=1,2}\left(\omega_j b^{\dag }_{j}b_{j}-g_{j}a^{\dag}a(b^{\dag}_{j}+b_{j})\right),\\ 
H_{int}&=J_m (e^{i\theta}b^{\dag }_{1}b_{2}+e^{-i\theta}b_{1}b^{\dag }_{2}),\\
H_{drive}&=i\sqrt{\kappa}\alpha^{\rm in}(a^{\dag}+a).  
\end{align}
where $H_{OM}$, $H_{int}$ and $H_{drive}$ are the optomechanical, phonon-hopping interaction  and the driving Hamiltonian, respectively. In Eq.(\ref{Hamil1}), $H_{\kappa}$ ($H_{\gamma}$) is the Hamiltonian capturing the optical (mechanical) dissipation. The variables $a$ and $b_{j}$ are the annihilation bosonic field operators describing the optical and mechanical resonators, respectively. The mechanical displacements $x_{j}$ are connected to the operators $b_{j}$ as $x_j=x_{ZPF}(b_j+b^{\dag }_{j})$, where $x_{ZPF}=\sqrt{\frac{\hslash}{2m\omega_j}}$ is the zero-point fluctuation amplitude of the mechanical resonator with $\omega_j$ the mechanical frequency of the $j^{th}$ resonator.  The frequency detuning between the driving ($\omega_p$) and the cavity ($\omega_{cav}$) is defined by $\Delta=\omega_p-\omega_{cav}$. The driving strength is defined by $\alpha^{\rm in}$ and $g_j$ is the optomechanical coupling. From the Heisenberg's equation $\dot{\mathcal{O}}=i[H,\mathcal{O}]+\mathcal{N}$, where $\mathcal{O}\equiv (a,b_j)$ and $\mathcal{N}\equiv (a^{in},b^{in}_j)$ the related noise operators, the quantum Langevin equations (QLEs)  of our system read,
\begin{equation}\label{Dyna}
\begin{cases}
\dot{a}&=\left(i\left(\Delta-\sum_{j=1,2}g_j (b^{\dag }_{j}+b_{j})\right)-\frac{\kappa}{2}\right)a +\sqrt{\kappa}\alpha^{in}+\sqrt{\kappa}a^{in},\\
\dot{b_1}&=-(i\omega_1 +\frac{\gamma_1}{2})b_1 -i J_m b_2 e^{i\theta}-ig_1 a^{\dag}a + \sqrt{\gamma_1}b^{in}_1,\\ 
\dot{b_2}&=-(i\omega_2 +\frac{\gamma_2}{2})b_2 -i J_m b_1 e^{-i\theta}-ig_2 a^{\dag}a + \sqrt{\gamma_2}b^{in}_2. 
\end{cases}
\end{equation}
These noise operators $\mathcal{N}\equiv (a^{in},b^{in}_j)$ have zero mean and are characterized by the following autocorrelation  functions \cite{Tchodimou.2017},
\begin{align}\label{noise}
\langle \mathcal{N}(t) \mathcal{N} ^{\dag} (t^{\prime}) \rangle &=(n_{\nu}+1)\delta(t-t^{\prime}),\\ 
\langle \mathcal{N}^{\dag}(t) \mathcal{N}(t^{\prime}) \rangle &=n_{\nu}\delta(t-t^{\prime}),
\end{align}
with $n_{\nu}\equiv (n_{th},n_{a})$, where  $n_{th}=\rm {exp(\frac{\hslash \omega_j}{k_B T}-1)^{-1}}$ and $n_{a}=\langle a^{in \dag} a^{in} \rangle$.

In this work, we are interested on the blue-sideband resonance where $\Delta=\omega_m$, with $\omega_m=\omega_1$. The sensitivity performance of our proposed sensor is investigated by computing the eigenvalues of the effective mechanical system which allows us to identify the EP. To achieve that, we need to linearize the QLEs given in Eq.(\ref{Dyna}) and then integrate the optical field out of the resulted system. The linearization process consists of splitting the field operators as $\mathcal{O}=\langle O \rangle +\delta \mathcal{O}$, where $\langle O \rangle$ is the coherent complex part of the operator and $\delta \mathcal{O}$ its related fluctuation. The linearized dynamics (See details in Appendix \ref{App.A}) can be treated by moving into another interaction picture by introducing the following slowly varying operators with tildes, $\delta a=\delta \tilde{a}e^{i\tilde{\Delta}t}$ and  $\delta b_j=\delta \tilde{b_j}e^{-i\omega_j t}$ where $\tilde{\Delta}$ is the effective detuning defined in Appendix \ref{App.A}.  In the limit of $\omega_j\gg(G_j,\kappa)$ with $G_j=g_j\langle a\rangle$, the rotating wave approximation can be invoked, and we can obtain the following equations: 
\begin{equation}\label{fluc}
   \delta \dot{\tilde{\mathcal{O}}}=-\rm{M}\delta \mathcal{\tilde{O}}+ \sqrt{\rm{K}} \delta\mathcal{\tilde{N}}
\end{equation}
where the matrix $\rm{M}$ is given by,
\begin{equation}\label{cor_mat}
\rm{M}=
\begin{pmatrix}
\frac{\kappa}{2} & iG_1 & iG_2\\
iG_1^{\ast} & \frac{\gamma_1}{2} & iJ_m e^{i\theta}\\
iG_2^{\ast} & iJ_m e^{-i\theta} & \frac{\gamma_2}{2}
\end{pmatrix},
\end{equation}
and the noise coefficients are $\rm{K}=(\kappa, \gamma_1, \gamma_2)^{T}$. Hereafter, and without loss of generality, we will assume that the optomechanical coupling strengths $G_1$ and $G_2$ are positive real numbers and will be set to be equal, i.e., $G_1=G_2=G$. Under the condition $\kappa\gg(G,\gamma_j)$ which is satisfied in our investigation, one can adiabatically eliminate the cavity field in Eq.(\ref{fluc}), and this leads us to the effective dynamical system for the two mechanical resonators (See details in Appendix \ref{App.A}), 
\begin{equation}\label{eff}
   i\partial_t \psi=\rm{H_{eff}}\psi,
\end{equation}
with $\psi=(\delta b_1, \delta b_2)$ and the effective Hamiltonian given by, 
\begin{equation}\label{eff_H}
\rm{H_{eff}}=
\begin{pmatrix}
\omega_1 -  \frac{i}{2}(\Gamma +\gamma_1) & J_m e^{i\theta}-i\frac{\Gamma}{2} \\
J_m e^{-i\theta}-i\frac{\Gamma}{2}  & \omega_2 -  \frac{i}{2}(\Gamma +\gamma_2)
\end{pmatrix},
\end{equation}
where the optomechanical induced damping is defined as $\Gamma=\frac{4G^2}{\kappa}$. The eigenvalues of the Hamiltonian given in Eq.(\ref{eff_H}) are,
\begin{equation}\label{Eig}
\lambda_{\pm}=\frac{1}{2}(\omega_1 +\omega_2)+\frac{i}{4}(\gamma_{eff}^1 +\gamma_{eff}^2) \pm \frac{\sigma}{4},
\end{equation}
where $\gamma_{eff}^j=\Gamma+\gamma_j$ and the quantity $\sigma$ is defined as,  
\begin{equation}\label{Sig}
\sigma=\sqrt{\chi-16iJ_m \Gamma \cos{\theta}},
\end{equation}
with $\chi=(2\Delta\omega+i\Delta\gamma)^2+16J_m^2-4\Gamma^2$, $\Delta\omega=\omega_1 - \omega_2$ and $\Delta\gamma=\gamma_{eff}^1 -\gamma_{eff}^2$. 
From these eigenvalues, the eigenfrequencies and eigendampings of the system are defined as being the real ($\omega_{\pm}=\Re{(\lambda_{\pm})}$) and imaginary ($\gamma_{\pm}=\Im{(\lambda_{\pm})}$) part of $\lambda_{\pm}$, respectively. At the EP, both these pairs of frequencies and dampings coalesce, i.e., $\omega_{-}=\omega_{+}$ and $\gamma_{-}=\gamma_{+}$, which means that $\sigma=0$. This condition on $\sigma$ can be fulfilled in our proposal by tuning the driving field or the phase $\theta$. For a proof of concept of the proposed sensor based on synthetic magnetism EP engineering, we assume in the next section that the two mechanical resonators are slightly  nondegenerated allowing us to not be limited by the constraints related to the microfabrication variability.   

\section{Sensitivity enhancement through synthetic magnetism} \label{sec:Sen}

The efficiency of our sensor sensitivity depends on the spectrum splitting at the EP once our system is perturbed. The perturbation can be a nanoparticule that has landed on the system or any mass deposition that is able to perturb the spectrum of our system. Therefore, we need to localize the EP before looking into our sensor feature. By assuming at first that $\gamma_{1}=\gamma_{2}=\gamma_m$ and that there is no direct coupling between the mechanical resonators ($J_m=0$) for instance, $\sigma$ reduces to, 
\begin{equation}\label{Sig0}
\sigma=2\sqrt{\Delta\omega^2-\Gamma^2},
\end{equation}
leading to an exceptional point at $\Gamma=\Delta\omega$, that is a result of an anti-PT symmetric character of our system coming from the dissipative coupling induced by the cavity \cite{Jiang_2021}. 
\begin{figure}[tbh]
  \begin{center}
  \resizebox{0.4\textwidth}{!}{
  \includegraphics{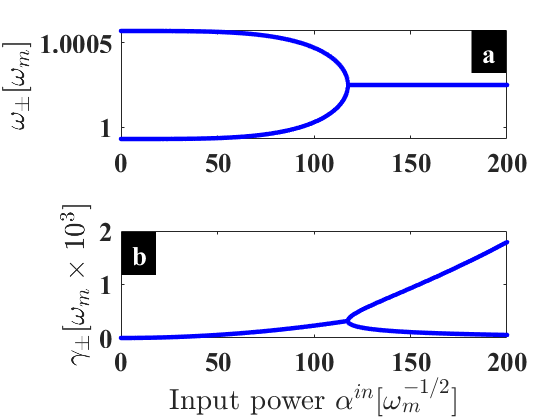}}
  \end{center}
  \caption{(a) Eigenfrequencies and (b) eigendampings of the system in the presence of the synthetic magnetism. The parameters used are $\omega_1=\omega_m$, $\omega_2=(1+5\times10^{-4})\omega_m$, $\kappa=7.3\times10^{-2}\omega_m$, $\Delta=\omega_m$, $g=1.077\times10^{-4}\omega_m$, $\gamma_1=1.077\times10^{-5}\omega_m$, $\gamma_2=\gamma_1$, $J_m=2\times10^{-4}\omega_m$ and $\theta=\frac{\pi}{2}$.}
  \label{fig:Fig2}
  \end{figure}

  \begin{figure}[tbh]
  \begin{center}
  \resizebox{0.4\textwidth}{!}{
  \includegraphics{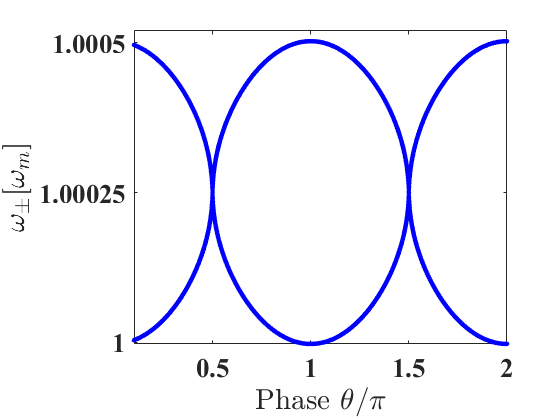}}
  \end{center}
  \caption{Eigenfrequencies versus $\theta$, where we can see the EPs engineered through the phase modulation at $\frac{\pi}{2}(2n+1)$. The parameters used are the same as before, and the driving field amplitude corresponds to the EP ($\alpha^{in}\sim117[\omega_m^{1/2}]$) shown in Fig. \ref{fig:Fig2}.}
  \label{fig:Fig3}
  \end{figure}
\begin{figure}[tbh]
  \begin{center}
  \resizebox{0.4\textwidth}{!}{
  \includegraphics{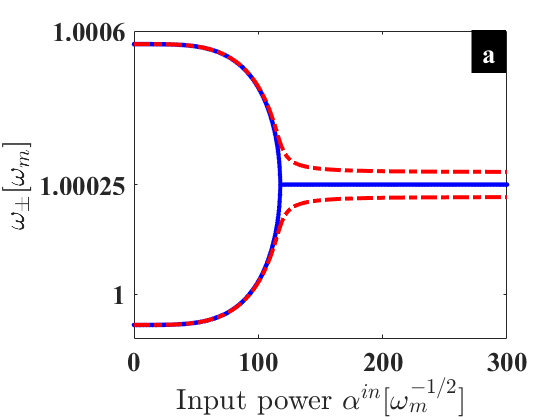}}
  \resizebox{0.4\textwidth}{!}{
  \includegraphics{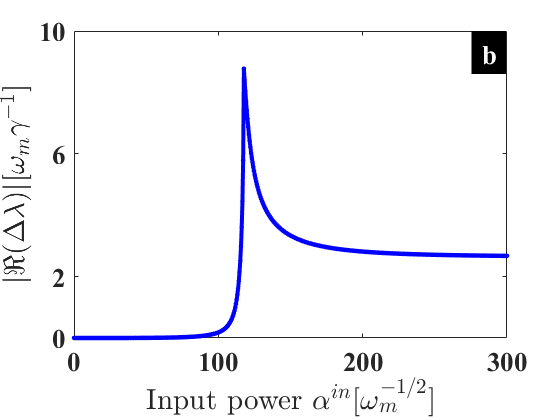}}
  \end{center}
  \caption{(a) Eigenfrequencies splitting and (b) the resulted splitting after a perturbation (see dashed curve in (a)). The parameters used are the same as in Fig. \ref{fig:Fig2}, and the perturbation strength is $\epsilon=0.2$.}
  \label{fig:Fig4}
  \end{figure}
\begin{figure}[tbh]
  \begin{center}
  \resizebox{0.4\textwidth}{!}{
  \includegraphics{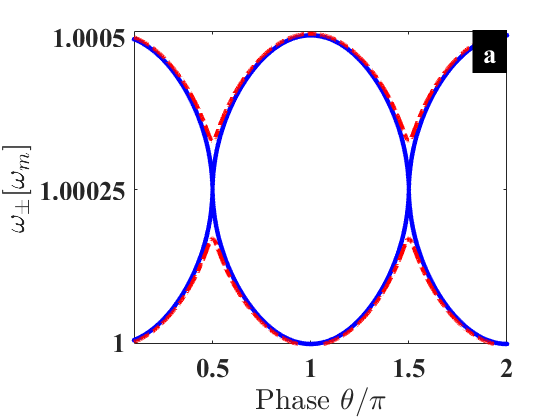}}
  \resizebox{0.4\textwidth}{!}{
  \includegraphics{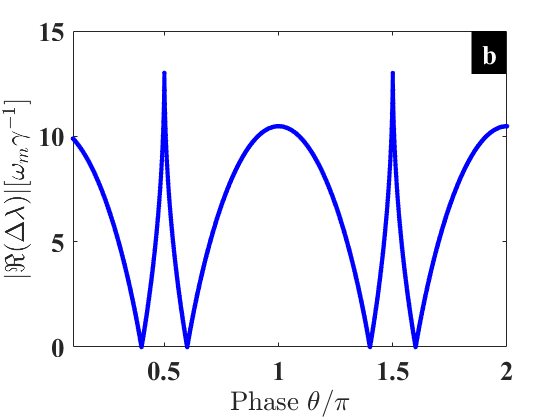}}
  \end{center}
  \caption{(a) Eigenfrequencies and (b) the resulted splitting versus $\theta$ after a perturbation ($\epsilon=0.5$). The parameters used are the same as in Fig. \ref{fig:Fig2}. The driving field amplitude of the dashed curve in Fig. \ref{fig:Fig5}a is $\alpha^{in}= 115[\omega_m^{1/2}]$, and it shows how the EP can be lifted when the field amplitude is not well tuned. (b) shows the possibility of multiple detection at the different EP.}
  \label{fig:Fig5}
  \end{figure}
\begin{figure}[tbh]
  \begin{center}
  \resizebox{0.4\textwidth}{!}{
  \includegraphics{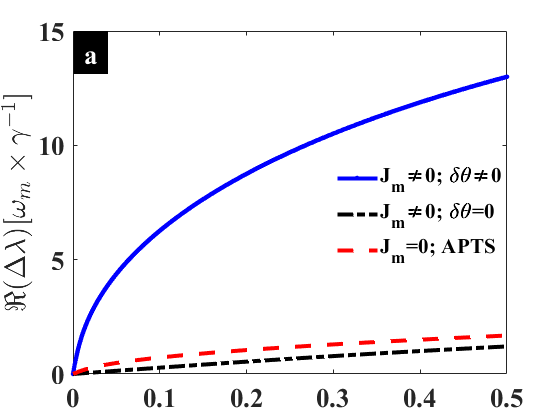}}
  \resizebox{0.4\textwidth}{!}{
  \includegraphics{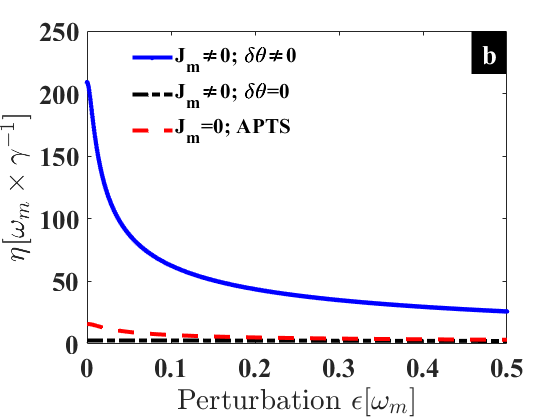}}
  \end{center}
  \caption{(a) Sensitivity at the exceptional point versus the
strength of the perturbation $\epsilon$. (b) Enhancement factor $\eta$ versus $\epsilon$. It can be
seen that for weak perturbation our proposed sensor performs better, and  this  performance decreases as the perturbation increases. Also, the sensor under synthetic magnetism is more efficient than the two  other cases (see dashed and dash-dotted curves in (a) and (b)).}
  \label{fig:Fig6}
  \end{figure}

By turning on the phonon-hopping coupling ($J_m\neq0$), this EP is shifted to $\Gamma=\sqrt{\Delta\omega^2+4J_m^2}$ providing that $\theta=\frac{\pi}{2}(2n+1)$ with $n$ being an integer. In Fig. \ref{fig:Fig2}, we have depicted the eigenfrequencies (Fig. \ref{fig:Fig2}a) and eigendamping Fig. \ref{fig:Fig2}b of our system for $J_m\neq0$, and they both coalesce at the EP. In order to show the $\theta$-dependence of the eigenvalues, we have sketched in Fig. \ref{fig:Fig3} the eigenfrequencies versus the phase and it can be seen that the EP happens at $\theta=\frac{\pi}{2}(2n+1)$.  To gain insight into our sensor feature, we assume that a mass (see the yellow dot in Fig. \ref{fig:Fig1}b for instance) has been deposited on the system. This mass will act as a perturbation on our system that will lift the degeneracy at the EP as shown in Fig. \ref{fig:Fig4}. Since our investigation is focused on the synthetic magnetism effect, we assume that this perturbation has mainly induced a phase-shift on our system. Therefore, the perturbed eigenvalues spectrum will depend on $\theta \pm\delta\theta$ instead, where  $\delta\theta$  is the phase shift resulted from the perturbation. Such a perturbation has led to the splitting shown in Fig. \ref{fig:Fig4}a where the perturbation has been introduced at the EP corresponding to $n=0$ ($\theta=\frac{\pi}{2}$). The related sensitivity is captured in Fig.\ref{fig:Fig4}b where the highest splitting shown by the peak at the EP reveals the efficiency of our proposed sensor. Moreover, the same sensitivity can be figured out when the EPs appearing for $n\neq0$ are perturbed as shown in Fig. \ref{fig:Fig5}, where Fig. \ref{fig:Fig5}a shows the splitting and Fig. \ref{fig:Fig5}b its related sensitivity. It can be seen that such a sensor can be used for detection at more than one EP, and such a feature may be interesting for multiple sensing scheme when several deposited masses have induced different perturbation strength. For quantification purpose, a perturbation strength of $\epsilon$ acts on the phase as $\theta=\pi/(2\pm\epsilon)\sim \frac{\pi}{2}(1\pm\epsilon)$ which lead the phase shift of $\delta\theta=\pm\epsilon\frac{\pi}{2}$.    
This phase shift acts on the eigenvalues given in Eq.(\ref{Eig}) as,
\begin{align}\label{Eigp}
\lambda_{\pm}^{\delta\theta}&=\frac{1}{2}(\omega_1 +\omega_2)+\frac{i}{4}(\gamma_{eff}^1 +\gamma_{eff}^2) \nonumber\\ &\pm \frac{1}{4}\sqrt{\chi-16iJ_m \Gamma (\cos{\theta}-\delta\theta\sin{\theta})}.    
\end{align}
We stress that this perturbation may also affects the optical dampings, but for the sake of a qualitative discussion and since the corresponding variations are very small, they were neglected in our analysis. To quantitatively give sense to the simulated results depicted in Figs. (\ref{fig:Fig4} and \ref{fig:Fig5}), we define the sensitivity as the absolute value of the frequency shift of an eigenfrequency from its reference signal, that is $\Re({\Delta\lambda_{\pm}})\equiv\Re(\lambda_{\pm}^{\delta\theta}-\lambda_{\pm})$. Owing to the high sensitivity at the EP (see the high peaks in Figs. (\ref{fig:Fig4} and \ref{fig:Fig5}), the related splitting of the eigenvalues can be deducted as (see Appendix \ref{App.B}),
\begin{equation}\label{SplitEP}
    \Delta\lambda_{\pm}^{EP}=\pm\frac{(1+i)}{2}\sqrt{2\delta\theta J_m \Gamma},
\end{equation}
which shows that there is no splitting at the EP for $\delta\theta=0$ as expected. Once the system is perturbed ($\delta\theta\neq0$), Eq.(\ref{SplitEP})
reveals that the eigenfrequencies experience the same splitting ($\Re(\Delta\lambda_{-})=\Re(\Delta\lambda_{+})$) as it can be seen in Figs. \ref{fig:Fig4}. Another feature revealed from Eq.(\ref{SplitEP}) is that the splitting at the EP scales as the square root of the strength of the perturbation ($\propto \delta\theta^{1/2}$), in stark contrast with the linear dependence for the conventional sensors \cite{Djorwe_2019}.  Owing to this complex square-root topology, the EP sensors perform better in detecting small mass deposition compared to the conventional ones as it can be seen in Fig. \ref{fig:Fig6} where both the sensitivity and the enhancement factor are shown. The enhancement factor is defined from Eq.(\ref{SplitEP}) as,
\begin{equation}\label{fac}
    \eta\equiv\left|\frac{\Re(\Delta\lambda^{E P})}{\delta\theta}\right|=\sqrt{\frac{J_m \Gamma}{2\delta\theta}}.
\end{equation}

For comparison purposes, we have considered two other cases where the phase $\theta$ is not perturbed. The first one is the anti PT-symmetric sensor case ($J_m=0$), where the perturbation will induce a change on the resonator quality factor ($\rm{Q_j}=\frac{\omega_j}{\gamma_j}$) rather than on the phase. The second case is when the two mechanical resonators are coupled ($J_m\neq0$), but the splitting is likewise induced by the quality factor perturbation. These two cases together with the phase shift perturbation are compared in Fig. \ref{fig:Fig6} for the sensitivity and enhancement factor respectively. In Fig. \ref{fig:Fig6}a, it can be seen a great improvement of the sensitivity when the phase is perturbed (full curve) as compared to the cases where the quality factor (the damping rate for instance) is perturbed. Moreover, regarding the cases where the quality factor is perturbed (dot-dashed and dashed curves), Fig. \ref{fig:Fig6}a reveals that the anti-PT-symmetry sensor (dot-dashed curve) has less sensitivity than the sensor operating at the EP engineered under synthetic magnetism condition which is not phase-perturbed (dashed curve). This feature reveals that phase perturbation has a profound impact on our sensor which in turn points out the sensing efficiency when operating under synthetic magnetism requirement. This sensing scheme based on EP which does not require a gain for amplification offers new opportunities to enhance sensing. Furthermore, Fig. \ref{fig:Fig6}b shows the enhancement sensing factor versus the perturbation strength. It can be seen a giant sensing enhancement under the phase-perturbation approach as aforementioned in  Fig.\ref{fig:Fig6}a. As the strength of the perturbation increases, one observes that the enhancement factor $\eta$ decreases. For strong-enough perturbation, $\eta$ evolves toward a limit bounded by the other two cases where the quality factor is perturbed. More importantly, the $\eta$ behaviour shows how greatly our sensor performs for weak perturbation strength, proving the efficiency of the EP sensor in detecting small particles, and diverse pollutants.         
  
\section{Conclusion} \label{sec:Concl} 

We have investigated sensor sensitivity based on exceptional points engineered under synthetic magnetism requirement. One advantage with this approach is to operate without a requirement of amplification gain, which substantially makes experimental tasks easier. Our system can be thought as an electromechanical (optomechanical) system where an electric (optical) field is driving two mechanical resonators which are mechanically coupled through a phase-dependent  phonon-hopping rate. We have shown that this phase induces series of EPs, which happen at the phase-matched condition of $\frac{\pi}{2}(2n +1)$. When there is no direct coupling between the two mechanical resonators, our system still hosts an EP owing to an anti-PT-symmetry feature coming form the dissipative coupling induced by the cavity. Owing to that, two perturbation schemes have been pointed out during our investigation, one related to the quality factor and the other one by acting on the phase-dependence. Our study has shown that the sensor perform better when the phase is perturbed rather than the quality factor. Moreover, this approach involving  synthetic magnetism allows multiple sensing scheme owing to the series of phase-induced exceptional points in the system. This work provides opportunities for sensitivity enhancement and sheds light on new platforms which can be used to develop efficient sensors.

\section*{Acknowledgments}
This work has been carried out under the Iso-Lomso Fellowship at Stellenbosch Institute for Advanced Study (STIAS), Wallenberg Research Centre at Stellenbosch University, Stellenbosch 7600, South Africa. 

P. Djorwe and S.G. Nana Engo thank the Ministry of Higher Education of Cameroon (MINESUP), for financial assistance within the framework of the ‘Research Modernization’ Allowances.

\appendix \label{App}

\section{Effective Hamiltonian} \label{App.A}
This section shows details of the dynamical equations from the Quantum Langevin Equations (QLEs) (see Eq.(\ref{Dyna}) in the main text) to the effective Hamiltonian given in Eq.(\ref{eff_H}). Starting with the QLEs, we linearize our system by splitting the bosonic operators $a$ and $b$ into their average part plus an amount of fluctuation as $a= \alpha +\delta a$ and $b_j=\beta_j +\delta b_j$, with $\alpha=\langle a \rangle$ and $\beta_j=\langle b_j \rangle$. The linearization process leads to the averaged dynamics, 
\begin{equation}\label{Av.dyna}
\begin{cases}
\dot{\alpha}&=\left(i\tilde{\Delta}- \frac{\kappa}{2}\right)\alpha+\sqrt{\kappa}\alpha^{in},\\
\dot{\beta_1}&=-(i\omega_1 +\frac{\gamma_1}{2})\beta_1 -i J_m e^{i\theta}\beta_2 -ig_1 |\alpha|^2,\\ 
\dot{\beta_2}&=-(i\omega_2 +\frac{\gamma_2}{2})\beta_2 -i J_m e^{-i\theta}\beta_1 -ig_2 |\alpha|^2, 
\end{cases}
\end{equation}
and to the fluctuation dyanmics,
\begin{equation}\label{Fluc.dyna}
\begin{cases}
  \delta \dot{a}&=\left(i\tilde{\Delta}- \frac{\kappa}{2}\right)\delta a-i\sum_jg_j(\delta b_j^{\dagger}+\delta b_j)\alpha+\sqrt{\kappa}a^{in},\\
\delta\dot{b_1}&=-(i\omega_1 +\frac{\gamma_1}{2})\delta b_1 -i J_m e^{i\theta}\delta b_2\\& -ig_1 (\alpha^{\dagger}\delta a+\alpha \delta a^{\dagger})+\sqrt{\gamma_1}b_1^{in},\\
\delta \dot{b_2}&=-(i\omega_2 +\frac{\gamma_2}{2})\delta b_2 -i J_m e^{-i\theta}\delta b_1\\&-ig_2 (\alpha^{\dagger}\delta a+\alpha \delta a^{\dagger})+\sqrt{\gamma_2}b_2^{in},
\end{cases}
\end{equation}

where the effective detuning has been defined as $\tilde{\Delta}=\Delta-2\sum_jg_j\Re{(\beta_j)}$. The stability of our system can be studied with the set of Eq.(\ref{Av.dyna}), while the dynamical fluctuation captured with Eq.(\ref{Fluc.dyna}) describes the behaviour of our system in the linearized regime. It can be treated by introducing the following slowly varying operators with tildes, $\delta a=\delta \tilde{a}e^{i\tilde{\Delta}t}$ and  $\delta b_j=\delta \tilde{b_j}e^{-i\omega_j t}$. This results to the following set of equations,
\begin{equation}\label{tilde.dyna}
\begin{cases}
\dot{\delta\tilde{a}}&=- \frac{\kappa}{2}\delta\tilde{a}-i\sum_jG_j(\delta \tilde{b_j}^{\dagger}e^{i(\omega_j-\tilde{\Delta})t}+\delta \tilde{b_j}e^{-i(\omega_j+\tilde{\Delta})t})\\&+\sqrt{\kappa}\tilde{a}^{in},\\
\dot{\delta\tilde{b_1}}&=-\frac{\gamma_1}{2}\delta\tilde{b_1} -i J_m \delta\tilde{b_2} e^{i\theta}e^{-i(\omega_2-\omega_1)t}\\&-ig_1 (\alpha^{\dagger}\delta \tilde{a}e^{i(\tilde{\Delta}+\omega_1)t}+\alpha \delta \tilde{a}^{\dagger}e^{-i(\tilde{\Delta}-\omega_1)t})+\sqrt{\gamma_1}\tilde{b_1}^{in},\\
\dot{\delta\tilde{b_2}}&=-\frac{\gamma_2}{2}\delta\tilde{b_2} -i J_m \delta\tilde{b_1} e^{-i\theta}e^{-i(\omega_1-\omega_2)t}\\&-ig_2 (\alpha^{\dagger}\delta \tilde{a}e^{i(\tilde{\Delta}+\omega_2)t}+\alpha \delta \tilde{a}^{\dagger}e^{-i(\tilde{\Delta}-\omega_2)t})+\sqrt{\gamma_2}\tilde{b_2}^{in}. 
\end{cases}
\end{equation}
For the sake of simplicity, we focus our attention during this investigation on the blue-sideband resonance $\tilde{\Delta}=\omega_j$. Under this assumption, the set of equations in Eq.(\ref{tilde.dyna}) reduces to, 
\begin{equation}\label{red.detu}
\begin{cases}
\dot{\delta\tilde{a}}&=- \frac{\kappa}{2}\delta\tilde{a}-i\sum_jG_j(\delta \tilde{b_j}^{\dagger}e^{2i\omega_j t}+\delta \tilde{b_j})+\sqrt{\kappa}\tilde{a}^{in},\\
\dot{\delta\tilde{b_1}}&=-\frac{\gamma_1}{2}\delta\tilde{b_1} -i J_m \delta\tilde{b_2} e^{i\theta}-ig_1 (\alpha^{\dagger}\delta \tilde{a}+\alpha \delta \tilde{a}^{\dagger}e^{-2i\tilde{\Delta}t})\\&+\sqrt{\gamma_1}\tilde{b_1}^{in},\\
\dot{\delta\tilde{b_2}}&=-\frac{\gamma_2}{2}\delta\tilde{b_2} -i J_m \delta\tilde{b_1} e^{-i\theta}-ig_2 (\alpha^{\dagger}\delta \tilde{a}+\alpha \delta \tilde{a}^{\dagger}e^{-2i\tilde{\Delta}t})\\&+\sqrt{\gamma_2}\tilde{b_2}^{in}. 
\end{cases}
\end{equation}
Under the limit of $\omega_j\gg(G_j,\kappa)$, we can invoke the rotating wave approximation by dropping the fast rotating terms, and  Eq.(\ref{red.detu}) further simplifies to,
\begin{equation}\label{fluc1}
\begin{cases}
\dot{\delta\tilde{a}}&=- \frac{\kappa}{2}\delta\tilde{a}-i\sum_jG_j\delta \tilde{b_j})+\sqrt{\kappa}\tilde{a}^{in},\\
\dot{\delta\tilde{b_1}}&=-\frac{\gamma_1}{2}\delta\tilde{b_1} -i J_m \delta\tilde{b_2} e^{i\theta}-iG^{\ast}_1 \delta \tilde{a}+\sqrt{\gamma_1}\tilde{b_1}^{in},\\
\dot{\delta\tilde{b_2}}&=-\frac{\gamma_2}{2}\delta\tilde{b_2} -i J_m \delta\tilde{b_1} e^{-i\theta}-iG^{\ast}_2 \delta \tilde{a}+\sqrt{\gamma_2}\tilde{b_2}^{in}, 
\end{cases}
\end{equation}
which is given in its compact form by Eq.(\ref{fluc}) in the main text. Under the  hierarchy of parameters $\kappa\gg(G,\gamma_j)$ which is fulfilled here, we can adiabatically eliminate the cavity field in Eq.(\ref{fluc1}) in order to get the mechanical effective system pointed out through Eq.(\ref{eff}).

\section{Perturbed eigenvalues and sensitivity} \label{App.B}
The phase perturbation of the eigenvalues given by Eq.(\ref{Eigp}) in the main text is explicitly written as,
\begin{equation}\label{Eigp1}
 \lambda_{\pm}^{\delta\theta}=\xi \pm \frac{1}{4}\sqrt{\chi-16iJ_m \Gamma \cos{(\theta +\delta\theta)}}.
\end{equation}
where we have set 
\begin{equation}
\xi=\frac{1}{2}(\omega_1 +\omega_2)+\frac{i}{4}(\gamma_{eff}^1 +\gamma_{eff}^2),
\end{equation}
and 
\begin{equation}
\chi=(2\Delta\omega+i\Delta\gamma)^2+16J_m^2-4\Gamma^2.
\end{equation}
As the perturbation $\delta\theta$ is small, one can assume
\begin{align}
\cos{(\theta +\delta\theta)}&=\cos(\theta)\cos(\delta\theta)-\sin(\theta)\sin(\delta\theta)\nonumber \\ 
&\sim\cos(\theta)-\delta\theta\sin(\theta),
\end{align}
which leads to,
\begin{equation}\label{Eigp2}
 \lambda_{\pm}^{\delta\theta}=\xi \pm \frac{1}{4}\sqrt{\sigma^2+16iJ_m \Gamma \delta\theta\sin(\theta)}. 
\end{equation}
The sensitivity of our system is evaluated as the difference between the perturbed and the nonperturbed spectra, i.e., $\Delta\lambda=\lambda_{\pm}^{\delta\theta}-\lambda_{\pm}$ as,
\begin{align}\label{Splt1}
 \Delta\lambda&=\lambda_{\pm}^{\delta\theta}-\lambda_{\pm}=\xi \pm \frac{1}{4}\sqrt{\sigma^2+16iJ_m \Gamma \delta\theta\sin(\theta)}-\left( \xi \pm \frac{\sigma}{4}\right)\nonumber\\
 &=\pm \frac{1}{4}\sqrt{\sigma^2+16iJ_m \Gamma \delta\theta\sin(\theta)}\mp\frac{\sigma}{4}. 
\end{align}
As we seek to evaluate this sensitivity at the EP, this means that the following requirements have to be considered, $\theta=\frac{\pi}{2}$ and $\sigma=0$. Therefore, the overall splitting at the EP yields,
\begin{equation}\label{Splt2}
 \Delta\lambda_{\pm}^{EP}=\pm \sqrt{i\delta\theta J_m \Gamma}.
\end{equation}
By using $\sqrt{i}=\frac{\sqrt{2}}{2}(1+i)$, Eq.(\ref{Splt2}) can be put in the form,
\begin{equation}\label{Splt3}
 \Delta\lambda_{\pm}^{EP}=\pm \frac{\sqrt{2}}{2}(1+i) \sqrt{\delta\theta J_m \Gamma},
\end{equation}
which clearly shows the square-root dependence to the perturbation $\delta\theta$ as expected. As the sensitivity is defined as $\Re({\Delta\lambda^{EP}})$, one finally deduces
\begin{equation}\label{Sens}
 \Delta\omega_{\pm}^{EP}=\Re{(\Delta\lambda_{\pm}^{EP})}=\pm \frac{\sqrt{2}}{2}\sqrt{\delta\theta J_m \Gamma}.
\end{equation}

\bibliography{Sensor}

\end{document}